# Density Function of Weighted Sum of Chi-Square Variables with Trigonometric Weights

**V. I. Egorov[a], B. V. Kryzhanovsky[a, *]**

[a]*Scientific Research Institute for System Analysis of the Russian Academy of Sciences, Moscow, 117312 Russia*

*\*e-mail: rvladegorov@rambler.ru*

**Abstract-**We have investigated a weighted chi-square distribution of the variable $\xi$ which is a weighted sum of squared normally distributed independent variables whose weights are cosines of angles $\varphi_k = 2\pi k / N$, where $k \in \{0,1,...,N-1\}$ and $N$ is the number of the freedom degrees. We have found the exact expression for the density function of this distribution and its approximation for large $N$. The distribution is compared with the Gaussian distribution.

**Keywords:** weighted chi-square distribution, quadratic forms, trigonometric weights, twice-degenerated weights, residue method.

## 1. INTRODUCTION

Let us consider the distribution of the random variable $\xi$:

$$\xi = \sum_{k=0}^{N-1} \lambda_k x_k^2, \tag{1}$$

where $x_k$ are independent normally distributed quantities with a zero mean and unit variance. The distribution (1) can be regarded as a weighted chi-square distribution with weights $\lambda_k$.

There are many researches on weighted sums of chi-square variables [1-6]. In the general case it is impossible to find a closed-form expression for the density function of this distribution. Different approximations using such techniques as Laguerre polynomial sequences [1, 2], expansions into gamma-distribution series [3, 4], etc., have been proposed. The reviews of the approximation methods are given in papers [5, 6]. The distribution of the form (1) is used in tests of fit in model validation [7], and other fields [8, 9].

Our paper concerns the case when weights $\lambda_k$ in the distribution (1) are defined as follows:

$$\lambda_k = 2 \cdot \cos \varphi_k , \quad \varphi_k = \frac{2\pi k}{N} , \quad k \in \{0,1,...,N-1\}. \tag{2}$$

The case has meaningful applications: researches [10, 11] demonstrate that eigenvalues of the matrix of the one-dimensional Ising model are defined by (2), and the energy distribution of the model is similar to the distribution of the variable (1).

## 2. GENERAL EXPRESSIONS

The density function of the variable (1) can be found by using the following expression

$$P(\xi) = \frac{1}{(2\pi)^{N/2}} \int_{-\infty}^{\infty} dx_1 \int_{-\infty}^{\infty} dx_2 ... \int_{-\infty}^{\infty} dx_N \; e^{-\frac{1}{2}\sum x_k^2} \delta\left(\xi - \sum_{k=0}^{N-1} \lambda_k x_k^2\right). \qquad (3)$$

The first moments of this distribution can be determined easily by simple integration of (3):

$$\langle \xi \rangle = \sum_{k=0}^{N-1} \lambda_k = 0, \quad \langle \xi^2 \rangle - \langle \xi \rangle^2 = 2\sum_{k=0}^{N-1} \lambda_k^2 = 4N. \qquad (4)$$

Replacing the delta function by its integral form and integrating on variables $x_k$, we get

$$P(\xi) = \frac{1}{2\pi} \int_{-\infty}^{\infty} \frac{e^{i\omega\xi}}{\prod_{k=0}^{N-1} \sqrt{1 + i2\omega\lambda_k}} d\omega. \qquad (5)$$

To be definite and to avoid loss of generality, we assume that the number of the degrees of freedom is even: $N = 2n + 2$. In such case $\lambda_k$ are twice degenerated variables ($\lambda_k = \lambda_{N-k}$) and there are only $n$ different weights $\Lambda_k = \lambda_k = \lambda_{N-k}$, $k = 1, 2, ..., n$. The exceptions are nondegenerate values $\lambda_0 = 2$ and $\lambda_{n+1} = -2$ which make singular not-a-pole points in the integrand. To simplify further calculations, let us exclude terms $\lambda_1 x_1^2$ and $\lambda_{n+1} x_{n+1}^2$ from the sum (1). Note that when $N$ is large enough, the contribution of these terms can be neglected. When $N$ is small, we can calculate their contribution at the final stage by using the convolution of independent distributions. In view of the foregoing assumptions, (5) takes the form:

$$P(\xi) = \frac{1}{2\pi} \int_{-\infty}^{\infty} \frac{e^{i\omega\xi}}{\prod_{m=1}^{n}(1 + i2\omega\Lambda_m)} d\omega. \qquad (6)$$

The integrand in (6) has $n$ first-order poles at points $\omega = i/2\Lambda_m$. When integrating (6) with $\xi > 0$, we should calculate residues at the poles in the upper half-plane $\text{Im}\,\omega > 0$. When $\xi < 0$, the poles in the lower half-plane $\text{Im}\,\omega < 0$ give a contribution in the integral. Considering that, the integration gives

$$P(\xi) = \begin{cases} \dfrac{1}{2} \displaystyle\sum_{\Lambda_m > 0} \dfrac{\Lambda_m^{n-2}}{\prod_{r \neq m}^{n}(\Lambda_m - \Lambda_r)} e^{-\frac{\xi}{2\Lambda_m}}, & \xi > 0 \\[2ex] -\dfrac{1}{2} \displaystyle\sum_{\Lambda_m < 0} \dfrac{\Lambda_m^{n-2}}{\prod_{r \neq m}^{n}(\Lambda_m - \Lambda_r)} e^{-\frac{\xi}{2\Lambda_m}}, & \xi < 0 \end{cases} \qquad (7)$$

Using the relationship proved in Appendix 1

$$\prod_{r\neq m}^{n}(\Lambda_m - \Lambda_r) = \frac{(-1)^m N}{4 - \Lambda_m^2} \tag{8}$$

we can finally rewrite (7) as

$$P(\xi) = \frac{1}{2N} \times \begin{cases} \sum_{\Lambda_m > 0} (-1)^m \Lambda_m^{n-2}(4 - \Lambda_m^2) e^{-\frac{\xi}{2\Lambda_m}}, & \xi > 0 \\ -\sum_{\Lambda_m < 0} (-1)^m \Lambda_m^{n-2}(4 - \Lambda_m^2) e^{-\frac{\xi}{2\Lambda_m}}, & \xi < 0 \end{cases} \tag{9}$$

Let us consider the general properties of the weighted distribution $P(\xi)$. At first we should note that for any set of different quantities $\Lambda_1, \Lambda_2, ..., \Lambda_n$, Euler-Lagrange equations [12] hold true:

$$\sum_{k=1}^{n} \frac{\Lambda_k^m}{\prod_{r\neq m}(\Lambda_k - \Lambda_r)} = \begin{cases} 1, & m = n-1 \\ 0, & 0 \leq m \leq n-2 \end{cases} \tag{10}$$

When $m = n-1$, the equation (10) corresponds to the normalization of the distribution $P(\xi)$. This may be seen by integrating (7). When $m = n-2$, (10) shows that is continuous at $\xi = 0$:

$$P(\xi \to 0_+) = P(\xi \to 0_-) \tag{11}$$

Indeed, it follows from (7) and (10) that

$$P(\xi \to 0_+) - P(\xi \to 0_-) = \sum_{k=1}^{n} \frac{\Lambda_k^{n-2}}{\prod_{r\neq m}(\Lambda_k - \Lambda_r)} = 0. \tag{12}$$

Similarly, (7) and (10) yield the relations:

$$\left.\frac{d^m P(\xi)}{d\xi^m}\right|_{\xi \to 0_+} = \left.\frac{d^m P(\xi)}{d\xi^m}\right|_{\xi \to 0_-}, \tag{13}$$

that means the continuity of derivatives of order $m \in [1, n-2]$ at point $\xi = 0$. Moreover, with a symmetric weight spectrum (when for each $\Lambda_k$ there is $\Lambda_r = -\Lambda_k$), it follows from (10) that

$$\left.\frac{dP(\xi)}{d\xi^k}\right|_{\xi = 0} = 0, \tag{14}$$

that is, the distribution $P(\xi)$ is symmetric about the point $\xi = 0$ and reaches the maximum at this point.

## 3. THE DENSITY FUNCTION WITH N >> 1

If $N \gg 1$, let us go from summation to integration over the variable $\varphi$ in (9):

$$P(\xi) = -\frac{1}{8\pi} \int_{-\pi/2}^{\pi/2} \frac{\sin^2 \varphi}{\cos^3 \varphi} \exp\left[\frac{1}{2} N f(\varphi)\right] d\varphi , \qquad (15)$$

where

$$f(\varphi) = i\varphi + \ln(2\cos\varphi) - \frac{\tilde{\xi}}{\cos\varphi}, \quad \tilde{\xi} = \frac{|\xi|}{2N} \qquad (16)$$

Since the distribution is symmetric around the central point $\xi = 0$, we introduce a convenient auxiliary $N$-independent variable $\tilde{\xi} \sim |\xi|$. Let us turn to the saddle-point method to evaluate integral (15). The equation for saddle point $f'(\varphi) = 0$ takes the following form:

$$t\left(t^3 + \tilde{\xi} t^2 + t - \tilde{\xi}\right) = 0, \quad t = e^{i\varphi} \qquad (17)$$

The point $t = 0$ is not a saddle one because $f''(t=0) = 0$. The root of equation $t^3 + \tilde{\xi} t^2 + t - \tilde{\xi} = 0$ is the saddle point. Let us introduce the necessary definitions:

$$p = 1 - \frac{\tilde{\xi}^2}{3}, \quad \alpha = \sqrt{\frac{|p|}{3}}, \quad Q = \frac{\tilde{\xi}}{3\alpha^3}\left[2 - \left(\frac{\tilde{\xi}}{3}\right)^2\right], \qquad (18)$$

Then for the saddle point we get

a) $$t = 2\alpha \cdot \operatorname{sh}\psi - \frac{1}{3}\tilde{\xi}, \quad \text{where} \quad \psi = \frac{1}{3}\operatorname{arcsh} Q \qquad (19)$$

when $p \geq 0$,

b) $$t = \begin{cases} \pm 2\alpha \cdot \operatorname{ch}\psi - \frac{1}{3}\tilde{\xi}, & \psi = \frac{1}{3}\operatorname{arcch}|Q|, \quad |Q| \geq 1 \\ 2\alpha \cdot \cos\psi - \frac{1}{3}\tilde{\xi}, & \psi = \frac{1}{3}\arccos Q, \quad |Q| < 1 \end{cases} \qquad (20)$$

when $p < 0$,

where the upper sign in (20) is taken when $Q > 0$, and the lower sign is chosen when $Q < 0$.

Then from (15) we get:

$$P(\xi) = P_0 e^{\frac{1}{2}Nf_0}, \quad \text{где} \quad f_0 = \ln(t^2 + 1) - \frac{2t\tilde{\xi}}{t^2 + 1} \qquad (21)$$

Here $f_0$ is the value of function $f(\varphi)$ at the saddle point, $t$ is determined by either (19) or (20), and the pre-exponential factor $P_0$ has the form:

$$P_0 = \frac{1}{2\sqrt{2\pi N(3t^2 + 2t\tilde{\xi} + 1)}} \left(\frac{t^2 - 1}{t^2 + 1}\right)^2 \qquad (22)$$

In the end let us evaluate the distribution function near its center, that is, when $|\xi| \ll N$. In this case we get from (19) that $t \approx \tilde{\xi}$ and the expression (21) takes the form

$$P(\xi) \approx \frac{1}{2\sqrt{2\pi N}} e^{-\frac{\xi^2}{8N}} \qquad (23)$$

As we see, in accordance with the central limit theorem, the distribution center follows the Gaussian distribution with variance $4N$.

## 4. ANALYSIS OF THE RESULTS

Figure 1 presents the density functions P($\xi$) drawn from (9), (21) and (23) for different $N$. To make the comparison more illustrative, the functions P($\xi$) are normalized by multiplying them by $N$. Since the distribution function is an even function, the graphs are drawn only for positive $\xi$. It is seen that with small $N$ the asymptotic expression (21) differs greatly from the exact distribution (9). For this reason, the series (9) should be used when $N < 500$. However, when $N$ grows, the asymptotic estimate (21) approaches the exact distribution function calculated by means of (9). The biggest estimation error of (21) is observed at the distribution function center near $\xi = 0$. At the same time the function (21) at its center follows the Gaussian distribution (23) closely. It should be noted that with large $N$ it becomes difficult to use the formula (9) because the terms of this alternating-sign series will be different in the digit exceeding the length of the mantissa of standard-type data. That is why the use of the asymptotic expression (21) makes more sense when $N \geq 500$.

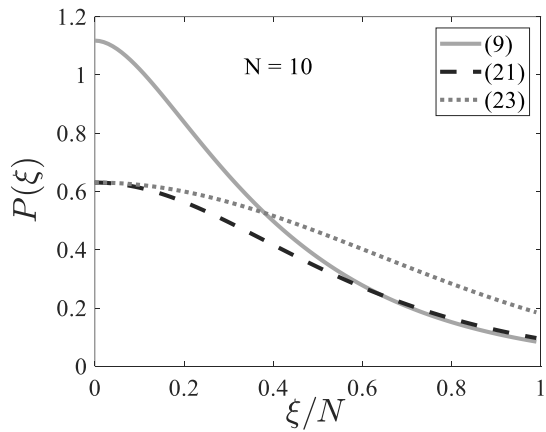

(a)

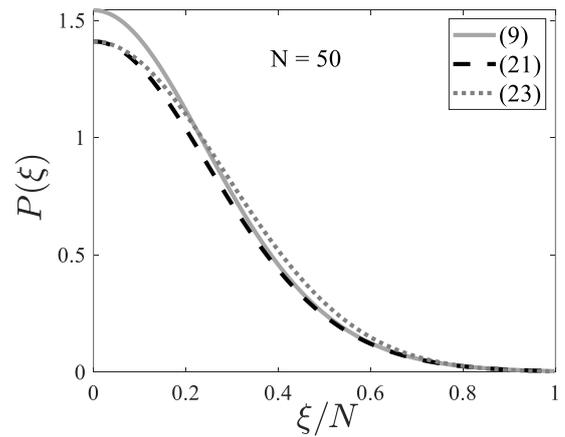

(б)

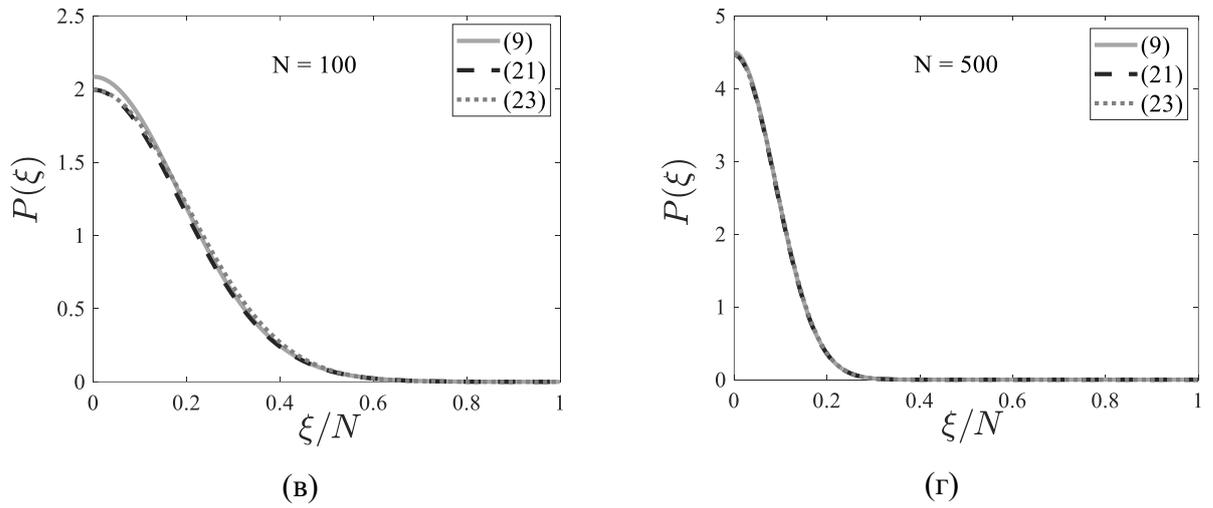

(в)  (г)

*Figure 1*. Distribution functions *P(ξ)* for *N* = 10 (a), 50 (b), 100 (c), 500 (d).

Figure 2 shows the logarithm of the distribution function. It is seen that the wing of the distribution (9) differs greatly from the Gaussian distribution (23). At the same time, the formula (21) provides a good representation of the distribution function (9) away from its center even with small N.

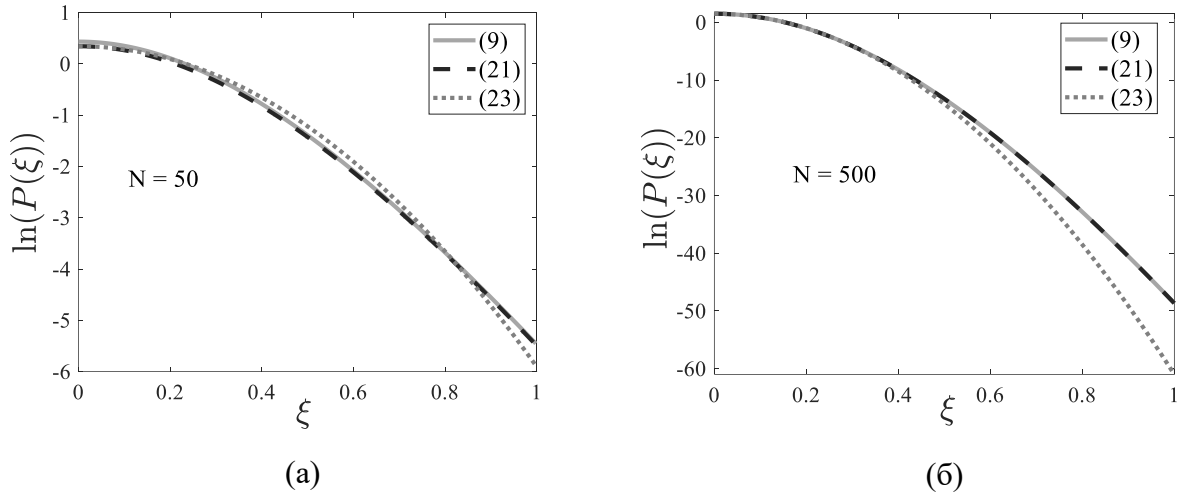

(a)  (б)

*Figure 2*. The logarithm of the distribution function *P(ξ)* for *N* = 50 (a), 500 (b).

## 5. CONCLUSION

We have obtained expression (9) for the distribution function of the random variable (1) whose weights are defined by the formula (2). The approximation (21), which is more suitable for large N, has been generated. The analysis shows that the distribution (9) is notably different from the Gaussian distribution.

Let us consider the product

$$\prod_{k=0}^{N-1}(\Lambda - \lambda_k) = 2^N \prod_{k=0}^{N-1}(\cos\varphi - \cos\varphi_k) = 2^{2N}\prod_{k=0}^{N-1}\sin\left(\frac{\varphi_k+\varphi}{2}\right)\cdot\prod_{k=0}^{N-1}\sin\left(\frac{\varphi_k-\varphi}{2}\right), \quad (A1)$$

where weights are defined by the expression (2) and a new variable $\Lambda = 2\cos\varphi$ is defined.

Using the well-known relation

$$2^{N-1}\prod_{k=0}^{N-1}\sin\left(x - \frac{\pi k}{N}\right) = \sin Nx \quad (A2)$$

we obtain the following expression for the product (A1):

$$\prod_{k=0}^{N-1}(\Lambda - \lambda_k) = -4\sin^2\frac{N\varphi}{2}. \quad (A3)$$

Extracting from (A3) the factors with nondegenerate weights $\Lambda - \lambda_0$ and $\Lambda - \lambda_{n+1}$, where $\lambda_0 = 2$ and $\lambda_{n+1} = -2$, and considering that other weights are twice degenerated, we obtain for the quantity $R = \prod_{m=1}^{n}(\Lambda - \Lambda_m)$:

$$R = \frac{1}{\sqrt{\Lambda^2 - 4}}\prod_{k=0}^{N-1}\sqrt{\Lambda - \lambda_k} = \frac{2}{\sqrt{4 - \Lambda^2}}\sin\frac{N\varphi}{2} \quad (A4)$$

From (A4) the expression (8) follows

$$\prod_{r\neq m}^{n}(\Lambda_m - \Lambda_r) = \lim_{\Lambda\to\Lambda_m}\frac{R}{\Lambda - \Lambda_m} = \frac{(-1)^m N}{4 - \Lambda_m^2} \quad (A5)$$

FUNDING

The research was supported by the state program FNEF-2022-0003 for the Research Institute of System Investigations of the Russian Academy of Sciences.

CONFLICT OF INTEREST

The authors declare that they have no conflicts of interest.